\documentstyle[11pt,aaspp4]{article}
 
\def\beq{\begin{equation}}
\def\enq{\end{equation}}
\def\bea{\begin{eqnarray}}
\def\ena{\end{eqnarray}}
\def\bec{\begin{center}}
\def\enc{\end{center}}
\def\etal{{\it et al.}}

\def\Ompi{(\Omega/4\pi)}

\def\Mesz{M\'esz\'aros~}

\def\eps{\epsilon}
\def\varep{\varepsilon}
\def\Fnum{F_{\nu_m}}
\def\Fnu{F_{\nu}}
\def\Inum{I'_{\nu_m}}
\def\num{\nu_m}
\def\siml{\lower4pt \hbox{$\buildrel < \over \sim$}}
\def\simg{\lower4pt \hbox{$\buildrel > \over \sim$}}

\begin{document}

\title{ GRB 990123: Reverse and Internal Shock Flashes\\
  and Late Afterglow Behavior}

\author{P. \Mesz$^{1,2}$ \& M.J. Rees$^3$} 
\noindent
$^1$Dpt. of Astronomy \& Astrophysics, Pennsylvania State University,
University Park, PA 16803 \\
$^2$Institute for Theoretical Physics, University of California, Santa Barbara,
CA 93106-4030\\ 
$^3$Institute of Astronomy, University of Cambridge, Madingley Road, Cambridge
CB3 0HA, U.K. 

\bec
{\it Submitted to MNRAS, 26 February 1999; accepted}
\enc


\begin{abstract}
The prompt $(t \siml 0.16$ days) light curve and initial
9-th magnitude optical flash from GRB 990123 can be attributed
to a reverse external shock, or possibly to internal shocks. 
We discuss the time decay laws and spectral slopes expected
under various dynamical regimes, and discuss the constraints imposed
on the model by the observations, arguing that they provide strongly
suggestive evidence for features beyond those in the simple standard model.
The longer term afterglow behavior is discussed in the context of the
forward shock, and it is argued that, if the steepening after three days is
due to a jet geometry, this is likely to be due to jet-edge effects, 
rather than sideways expansion.
\end{abstract}


\section{Introduction}

The observations of GRB 990123 (Akerlof, etal, 1999b; Kouveliotou, etal 1999;
Kulkarni et. al 1999) not only pose constraints on the amount of gamma-ray
beaming needed from a stellar mass progenitor in the absence of lensing,
but also provide an interesting test of the canonical fireball shock
afterglow model. A simultaneous optical flash of 9-th magnitude from
a burst at cosmological redshifts was discussed more than two years ago 
by \Mesz \& Rees 1997 (models a2, a3), from the reverse shock that acompanies 
the blast wave.  This early optical flash is expected to start at the time of 
the gamma-ray trigger, and to decay faster than the better known radiation from 
the forward blast wave, which starts out weaker but dominates the longer 
duration optical afterglow. 
A similar prediction based on the reverse shock was made in the light of 
more recent studies by Sari \& Piran (1999a), and more specifically discussed 
by them in the context of the observations of GRB 990123 (Sari \& Piran 1999b).  

A different origin for a simultaneous optical flash is possible from
internal shocks (\Mesz \& Rees, 1997, e.g. model b2 of that paper).  
Internal shock optical flashes are of additional interest because, 
as pointed out by Fenimore et al (1999), the gamma-ray light curve of GRB 990123 
(as well as those of several other bursts) appears to be incompatible with the 
gamma-rays coming from a single external shock, since the gamma-ray pulses in 
the second half of the burst are not appreciably longer than in the first half.
It is, of course, possible that the gamma-rays arise in internal shocks, 
which are thought to be exempt from such problems, while the optical 
afterglows may arise from the external and the associated reverse shocks. 
However, in the light of the need for internal shocks, it is interesting 
to investigate the implications of the early afterglow observations at 
various wavelengths including optical, for both external and internal shocks.

We also discuss the longer term behavior of the afterglow, and the
causes for the optical light curve flattening after 0.16 days.
The likely interpretation of the light curve after this time is that it
is due to the forward shock or blast wave. The discrepancy pointed out between 
the observed time decay slope and the spectral index within the context of the
simple standard model can be resolved by invoking the simplest realistic
extensions to this model, and we discuss several specific possibilities (\S 4).
We also indicate that the steepening of the light curve to $\propto t^{-1.8}$ 
after about three days, if due to a jet geometry, is likelier to be due to the 
effects of beginning to see the edge of the jet. This effect occurs before,
and its effects fit the steepening better, than the alternative sideways 
expansion interpretation (\S 5).

\section{Optical flash from reverse external shocks}

The reverse shock acompanying the forward blast wave gives the right
magnitude prompt optical flash with reasonable energy requirements
of no more than a few $10^{53}$ erg isotropic (\Mesz \& Rees 1997).
The time decay constants calculated in that paper for models a2, a3 were 
affected by an error, which we correct here; we consider also a more
generic prescriptin for the dynamics and the magnetic field behavior.

For a general evolution of the bulk Lorentz factor $\Gamma \propto r^{-g}$
with radius, the radius and observer time $t$ are related through $r\sim
ct\Gamma^2$, or
\beq
\Gamma\propto r^{-g}\propto t^{-g/(1+2g)}~~,~~ r\propto t^{1/(1+2g)}~.
\enq
In the usual Blandford-McKee (1976) impulsive solutions, the ``adiabatic"
case is $g=3/2,~\Gamma\propto r^{-3/2},~r\propto t^{1/4}$; the ``radiative" 
case is $g=3,~\Gamma\propto r^{-3},~r\propto t^{1/7}$  and in the similarity 
limit $g=7/2,~\Gamma\propto t^{-7/2},~r\propto t^{1/8}$. More general values 
of $g$ occur if the injection is non-uniform (Rees \& \Mesz 1998), anisotropic 
or the external medium is inhomogeneous (\Mesz, Rees \& Wijers 1998). Strictly
speaking $\Gamma$ is the bulk Lorentz factor of the forward shocked material,
and may be used also for the contact discontinuity. It is only a very 
rough approximation for the Lorentz factor of the reverse shock. Assuming
the latter approximation is valid, the comoving width, volume, and particle 
density of the ejecta, after it has been traversed by the reverse shock, 
evolve with
\beq
\Delta R \sim r/\Gamma \propto r^{1+g} ~~,~~ V' \sim {n'}_{ej}^{-1} \propto
 r^2 \Delta R \propto r^{3+g}.
\enq
We consider, in the reverse shocked gas, two possibilities for the comoving
magnetic field evolution: one is flux-freezing, $B'\propto V'^{-2/3}$,
and the other is that the comoving field in the reverse shocked gas 
remains in pressure quilibrium with the forward shocked gas, $B'\propto
\Gamma$. For flux-freezing (pressure eqiulibrium) we have then
\beq
B'\propto r^{-(6+2g)/3}~~,~~(\hbox{or} \propto r^{-g})~.
\enq
The energy density $\varep'$ and the electron random Lorentz factor $\gamma$ 
in the reverse shocked gas follow from $\varep'\propto V'^{-4/3}\propto
r^{-(12+4g)/3}$, (or $\propto \Gamma^2 \propto r^{-2g})$ and 
$\gamma \propto \varep' / n'_{ej} \propto r^{-(3+g)/3}$ ($\propto r^{3-g}$).
The synchrotron peak energy in the observer frame is then
\bea
\nu_m \propto \Gamma  B' \gamma^2 
& \propto r^{-(12+7g)/3} \propto t^{-(12+7g)/(3+6g)}~~\nonumber \\
(& \propto r^{6-4g} \propto t^{(6-4g)/(1+2g)} )~,
\ena
in these two magnetic field cases. Considering for simplicity the case where 
the electron cooling time is long compared to the dynamic expansion
time, the comoving synchrotron intensity at the peak frequency is
$\Inum \propto n'_{ej} B' \Delta R \propto r^{-(12+2g)/3}$ 
($\propto r^{-(2+g)}$), and the observer-frame flux is
\bea
\Fnum \propto t^2 \Gamma^5 I'_{\nu_m} 
& \propto r^{-(6+5g)/3}\propto t^{-(6+5g)/(3+6g)}~~ \nonumber \\
(& \propto r^{-2g}\propto t^{-2g/(1+2g)} )~.
\ena
For a photon energy spectral index $\beta$ ($\Fnu \propto \nu^\beta$) the 
spectral flux at a given frequency (e.g. optical) expected from the reverse 
shocked gas is then
\bea
\Fnu \sim & F_{\nu_m}.(\nu/\nu_m )^\beta \propto \Fnum \num^{-\beta}
 ~~~~~~~~~~~~~~~~~~~~~~~~~~~~~\nonumber\\ 
 & \propto r^{-[6-12\beta +g(5-7\beta )]/3}
    \propto t^{-[6-12\beta +g(5-7\beta)]/(3+6g)} \nonumber \\
(& \propto r^{-[2g(1-2\beta)+6\beta]}
  \propto t^{-[2g(1-2\beta)+6\beta]/(1+2g)} )~,
\ena
where $\beta=1/3$ below $\num$ and $\beta=-(p-1)/2$ above $\num$ in
synchrotron radiation.
We have assumed here that the cooling frequency is above the peak frequency,
the conditions for the latter appearing to be satisfied in GRB 990123, 
as pointed out by Sari \& Piran 1999b. Under the flux-freezing field behavior,
for an adiabatic case $g=3/2$ and an
electron index $p=2.5$, the photon spectral index above $\num$ is $\beta=-3/4$ 
and we have $\Fnu \propto t^{-81/32} \sim t^{-2.5}$, whereas for $p=2$ one 
would have $\beta=-1/2$ and $\Fnu \propto t^{-33/16} \sim t^{-2}$ 
(while for the similarity case $g=7/2$, $p=5/2$ we get, as Sari \& Piran 1999b,
$\Fnu \propto t^{-411/192} \sim t^{-2.1}$).
Under the pressure equilibrium field behavior, for $g=3/2$, and arbitrary 
$p$, we have $\Fnu\propto t^{-3/4}$ (this is a degenerate case where
$\num$ is constant), whereas for $g=7/2$, $p=5/2$ we have $\Fnu\propto 
t^{-13/8} \sim t^{-1.6}$.
The ROTSE observations (Akerlof etal, 1999) give an approximate dependence 
$\propto t^{-2}$ for about 600 seconds), which is in rough agreement with
the flux-freezing value in either the adiabatic $p=2$ or similarity $p=5/2$
cases, which cannot be distinguished without spectral information during 
the ROTSE observations.  A different analysis of the earliest data on GRB 990123
(Fruchter \etal 1999) gives the ROTSE slope as $\propto t^{-1.6}$, which is 
close to the pressure equilibrium similarity solution for $p=5/2$.

However, note that a decay $\propto t^{-2}$ (or $\propto t^{-1.6}$) can also
be obtained in more generic situations than the above, and in particular
the spectral slope need not be the only constraint on the decay index, except 
in the simplest, 
homogeneous external medium, or single initial $\Gamma$ (impulsive) model. 
For an impulsive injection in an inhomogeneous external medium, e.g. 
$\rho_{ext}\propto r^{-d}$ one expects in the adiabatic limit $\Gamma \propto
r^{-g}$ with $g = (3-d)/2$ (\Mesz, Rees \& Wijers 1998). For a more realistic
non-uniform injection situation, one expects a range of initial $\Gamma$, 
and in particular for a power-law distribution of $\Gamma$ in the ejecta 
(Rees \& \Mesz, 1998) where the mass injection rate $M(>\Gamma) \propto 
\Gamma^{-s}$, the effective resulting $\Gamma \propto r^{-g}$ with
$g=3/(2+s)$. The reverse shock relation (6) between photon spectral slope and 
dynamics could therefore, in the simplest ``post-standard" extensions of the 
model, depend e.g. on the external density profile, or the mass-energy 
injection dependence. 

\section{ Optical flashes from internal shocks}

Internal shocks, just as external shocks, should have a low energy 
portion or tail of the spectrum which extends into the optical, 
starting at the GRB trigger (except possibly in rare cases where the 
self-absorption frequency extends to the optical). For the same total energy,
the flux from the internal shock optical flash is about 2 orders of magnitude 
weaker than from the reverse external shock (\Mesz \& Rees 1997); however,
for beaming factors $\sim 10^{-2}$ as widely considered, they could lead
to flashes as bright as 9-th magnitude at $z\sim 1$. The simple ``standard
internal shock model" (Rees \& \Mesz, 1994) considers a wind of duration
$t_w \sim t_{burst}$ with $< \Gamma > \sim $ constant and fluctuations 
$\Delta\Gamma\sim \Gamma$ over times $t_v < t_w$ producing shocks at
a distance $r_{sh}\sim c t_v <\Gamma >^2$ for $t \siml t_w$.
1) The most straighforward GRB internal shock (e.g. model b1 of \Mesz \& 
Rees 1997) assumes that the gamma-ray break 
observed around $\sim 100$ keV is due to the synchrotron peak energy.
The magnetic field required in the internal shocks is not far below 
equipartition, and this leads to a very short radiative cooling time compared 
to the expansion time. If the shocks arise from discrete, sharp-edged shells,
there would be a very sudden drop of the light curve at all energies (including
optical) after internal shocks stop. However, a more realistic situation 
probably involves smoothly modulated shells, where the shocks weaken gradually
after reaching a peak strength. A variety of plausible distributions would 
then lead to a power law decay of the optical light.\\
2) An alternative internal shock model, which would lead to long decay even
in the absence of a gradual fading of the shocks, is obtained in the case
where the synchrotron break is at optical energies, and the gamma-ray break is 
produced by IC-scattered synchrotron photons (model b2 of \Mesz \& Rees 1997, 
Papathanassiou and \Mesz, 1996).  Between the time at which internal shocks (or 
gamma-rays) stop and the time when deceleration by the external medium starts
($t_{dec} \sim 500 (E_{54.6}/\Ompi n_{ext})^{1/3} \Gamma_{300}^{-8/3}$ sec),
the average bulk Lorentz factor remains approximately unchanged, and the 
dynamics are described by $\Gamma \propto t^0 ,~ r\propto t $. Hence, the 
comoving width, volume, and particle density of the ejecta evolve as
\beq
\Delta R \sim r/\Gamma \propto t ~~,~~ V' \sim {n'}_{ej}^{-1} \sim 
r^2\Delta R \propto t^3,
\enq
and the comoving energy density and electron random Lorentz factor will be
\beq
\varep' \sim n' \gamma \propto V'^{-4/3} \propto t^{-4} ~~,~~
\gamma \sim \varep' / n' \propto t^{-1}~.
\enq
If the magnetic field is not dynamically dominant, two cases are \\
a) $ B'\propto V'^{-2/3} \propto t^{-2}$, if the field is random, and \\
b) $B'\propto B \propto r^{-1}\propto t^{-1}$, if the field is mainly 
transverse (e.g. inefficient reconnection). Cases a (b) lead to an 
observer-frame synchrotron peak flux 
\beq
\nu_m \propto \Gamma . B' . \gamma^2 \propto t^{-4}~~(\propto t^{-3})
\enq
(The first dependence assumes random fields , and the second dependence is for
transverse fields). The comoving intensity is then
$I'_{\nu_m} \propto n'_{ej} B' \Delta R \propto t^{-4} ~~(\propto t^{-3})$
and
\beq
F_{\nu_m} \propto t^2 \Gamma^5 I'_{\nu_m} \propto t^{-2} ~~(\propto t^{-1} ),
\enq
while the flux at a fixed frequency is
\beq
\Fnu \sim \Fnum \num^{(p-1)/2} \sim t^{-2p}~~ (\sim t^{(1-3p)/2}),
\enq
where one can verify that at the internal shock radius and above, the 
optical electrons are in the adiabatic regime, for a wind equipartition
parameter $\eps_B=10^{-6}$ and $\gamma\sim 300$.
Thus for $p=2$, $\Fnu \propto t^{-4 }~~ (\propto t^{-5/2})$

For a magnetically dominated outflow, an interesting third case is \\
c) $B'\propto B \propto t^{-1}$, as in (b) above, but now one might
expect the comoving volume and particle density to be dominated by the
field evolution (via $B' \propto t^{-1} \propto V'^{-2/3}$), so 
\beq
V' \propto {n'}_{ej}^{-1} \propto t^{3/2} ~~,~~ 
 \Delta R \sim V'/r^2 \propto t^{-1/2},
\enq
\beq
\varep' \sim n' \gamma \propto V'^{-4/3} \propto t^{-2} ~~,~~
\gamma \sim \varep' / n' \propto t^{-1/2},
\enq
\beq
\nu_m \propto \Gamma . B' . \gamma^2 \propto t^{-2}
\enq
the comoving intensity is $I'_{\nu_m} \propto n'_{ej} B' \Delta R \propto 
t^{-3}$ and
\beq
F_{\nu_m} \propto t^2 \Gamma^5 I'_{\nu_m} \propto t^{-1}
\enq\beq
\Fnu \sim \Fnum \num^{(p-1)/2} \sim t^{-p}.
\enq
Adiabatic conditions prevail for $\eps_B\sim 1$ and, for example, $\Gamma 
\simg 300$, $\gamma \sim 100$. For $p=5/2$ this gives $\Fnu\propto 
t^{-5/2}$, and for $p=2$ it gives $\Fnu\propto t^{-2}$. 

For the simple internal shock model where the average $\Gamma$ is constant,
model (a) is too steep, but model (b) and (c) could fit the observations, 
with a flat enough electron power law ($p\sim 2$). However, in a realistic
model the average $\Gamma$ of the wind producing internal shocks could vary,
and even for $p> 2$, a decay $t^{-2}$ or $t^{-1.6}$ could be the result, 
e.g., of an average $\Gamma$ which increases in time. 

\section{The external blast wave}

The standard long-term (as opposed to short-term) afterglow is attributed to 
the external blast wave (forward shock) evolution, whose bulk Lorentz factor 
is described by the same equations (1), and the comoving width and volume of the forward
shocked gas by equations (2). The comoving density of the forward shocked gas 
is, however, $n'\sim n_{ext}\Gamma$ and the comoving field is assumed to be 
some fraction of the equipartition value, $B'\propto \varep_B^{1/2}\Gamma$, 
while the shocked electron random Lorenzt factor is $\gamma \sim \varep_e 
\Gamma$. As in \Mesz \& Rees 1997 (model a1), $\num\propto r^{-4g}\propto 
t^{-4g/(1+2g)}$, for adiabatic electrons $\Inum\propto n' B' \Delta R \propto
r^{1-g}$, $\Fnum\propto t^2\Gamma^5\Inum\propto r^{3-2g} \propto 
t^{(3-2g)/(1+2g)}$, and for a spectrum $\propto \nu^\beta$ we have
\beq
\Fnu\propto r^{3-2g(1-2\beta)}\propto t^{[3-2g(1-2\beta)]/(1+2g)}.
\enq
Thus in the (impulsive) ``standard model" a simple relation is expected 
between the time decay index 
$\alpha$  and spectral slope $\beta$, namely $\alpha= [3-2g(1-2\beta)]/(1+2g)$. 
E.g. for the impulsive adiabatic case $g=3/2$ and one would expects 
$\alpha=(3/2)\beta$ with $\beta=1/3,~-(p-1)/2$ for synchrotron, or their 
equivalent for the radiative $g=3$ or similarity $g=7/2$ values. 
Such a simple one-parameter relation between $\alpha$ and $\beta$ does not
appear to hold for the second stage of GRB 990123, where $F_{opt}\propto 
t^{-1.1}$ (Kulkarni, \etal, 1999). 
This, in our view, is a strong indication that ``post-standard" features 
are present, e.g. a non-homogeneous external medium or an anisotropic outflow 
(\Mesz, Rees \& Wijers, 1998), or non-uniform injection (Rees \& \Mesz, 1998).
From an observed time decay $F_{opt} \propto t^{-\alpha}$ and an observed 
spectral slope $F_\nu \propto \nu^\beta$ one can then work backwards to get 
the effective value of $g$ implied by the equations (6) above, and this in turn 
implies a value of $d=3-2g$ or $s=(3/g)-2$, thus providing information about
the external medium or the injection mechanism. While non-unique, such examples 
indicate that the time decay index is likely to depend on parameters other 
than the spectral slope.

\section{A Jet Geometry? Edge vs. Expansion Effects}

The temporal decay index after three days steepens from about $t^{-1.1}$
to $t^{-1.8}$ (Kulkarni, \etal 199; Fruchter, \etal 1999), and this
steepening can be attributed to a collimated
outflow, whose effects become noticeable after $\Gamma$
drops sufficiently. If one assumes (cf Kulkarni, \etal 1999) that the steepening
is caused by  sideways expansion of the decelerating  jet (Rhoads, 1997) 
one expects a large steepening from $t^{-1.1}$ to $t^{-p}$, where $p$ is
electron index, typically $p=2.5$, so the change expected would be more than
one power of $t$. However, the edge of the jet begins to be seen when $\Gamma$
drops below the inverse jet opening angle $1/\theta_j$. This occurs well before
sideways expansion starts (Panaitescu \& \Mesz, 1998):
the latter is unimportant until the expansion is almost  non-relativistic. 
 So long as  $\Gamma >\theta_j^{-1}$, the emission we receive is the same as
from a spherically-symmetric source, and  the effective transverse area is 
$A \sim (r_\parallel/\Gamma)^2 \propto t^2\Gamma^2$; on the other hand, when
$\Gamma <\theta_j^{-1}$ the dependence is 
 $A\sim (r_\parallel \theta_j)^2 \propto t^2 \Gamma^4$. Note that even
after the edge of the jet becomes visible, the outflow is still essentially
radial and $\Gamma$ continues to decay as a
power law as before (until sideways expansion sets in). These two
additional powers of $\Gamma$ (e.g. in adiabatic expansion, $\Gamma \propto
r^{-3/2}\propto t^{-3/8}$) imply a steepening by $t^{-3/4}$, which matches quite
well the observed steepening by a power of about $t^{-0.7}$.
There are of course several other plausible causes for steepening, as discussed
by Kulkarni \etal (1999), \Mesz, Rees \& Wijers 1998, Rhoads 1997 and others.
However, the agreement between the observed change in the decay slope
and that expected from seeing the edge could well be a signature of the
detection of a jet.

\section{Discussion}

The conclusions that may be drawn from the above are that bright
optical flashes, starting at the time of the gamma-ray trigger
and extending into the early afterglow, are a robust prediction
of the simplest afterglow models, as discussed in \Mesz \& Rees, 1997,
Sari \& Piran, 1999a, 1999b. A 9-th magnitude optical flash from a GRB at
redshift $z \sim 1$ can arise from a reverse external shock with total
energy $E_{53}\Ompi^{-1} \sim 1$, or from internal shocks  with 
$E_{53}(\Omega/ 10^{-2} 4\pi)^{-1} (t_\gamma /10^2 s)^{-1} \sim 1$ 
(\Mesz \& Rees 1997), where $\Omega$ is the solid angle into which the
radiation is collimated, and $t_\gamma$ is the GRB duration.
As known from $\log N-\log P$ fits (e.g. Krumholz etal, 1999), in 
essentially all cosmological models the luminosity function must be
broad, with a range of $E$ which can span upwards of 2-3 orders of magnitude.
This is compatible with the fact that GRB 990123 is in the top 1\%
of the BATSE brightness distribution (Kouveliotou etal, 1999), and also
with the previous non-detection of similarly bright optical flashes.

A steeper time-decay law is expected for the early optical flash from 
reverse or internal shocks, compared to that expected from the forward
shock. The decays calculated in \Mesz \& Rees 1997 are too steep, compared
to the observed $t^{-2}$ behavior (Akerlof etal 1999), indicating the need
for investigating different assumptions for the magnetic field behavior.
In the present paper we find that an early time-decay $\propto t^{-2}$
is naturally explained in the standard (single-$\Gamma$) afterglow model 
either by reverse shocks, where the field is in pressure equilibrium with the
forward shock or is frozen-in, or by internal shocks in a magnetically 
dominated outflow. However, such decays may also be easily achieved in more 
realistic afterglow models without severe restrictions on the electron index, 
e.g. for bursts in an inhomogeneous external medium, or characterized by a 
power law range of $\Gamma$.

From the fact that the inferred isotropic equivalent gamma-ray energy is
$4\times 10^{54}$ ergs (Kouveliotou etal, 1999), one infers the need
for a collimation of the gamma-rays by at least $10^{-1}$, if not $10^{-2}$, 
for any stellar-mass source. For an external reverse shock origin, from the
energy alone the optical radiation need not be collimated at all; however, 
since in the observer frame the Lorentz factor of the reverse shock is 
initially close to that of the blast wave, the initial optical and gamma-ray 
collimation should be approximately the same (except, as may be the case, if 
the jet has an anisotropic $\Gamma$). For an internal shock origin of the 
optical flash, there are stronger grounds for expecting the beaming of the 
optical to be the same initially as for the gamma-rays. 

The flattening of the light curve after 0.16 days to $\propto t^{-1.1}$ is well 
explained by the optical light expected from the forward shock, conforming to 
the standard interpretation of the long-term behavior of afterglows. 
The discrepancy pointed out Kulkarni \etal 1999 between the standard model 
prediction and the observed relation between the time decay index and the 
spectral slope is, as pointed out in \S 4, most likely to be an indication of 
departures from the simple standard model, e.g. an inhomogeneous external 
medium, non-uniform injection, or anisotropies. 
The renewed steepening of the optical light curve after about three days 
to $\propto t^{-1.8}$ (Kulkarni, \etal 1999, Fruchter, \etal 1999) is, 
as argued in \S 5, likely to be due to seeing the edge of a jet,  which
gives a better fit for the magnitude of the change of slope than expected 
from sideways expansion, and also should occur well before the latter begins.

There are two main arguments that militate in favor of the optical flash 
observed in GRB 990123 being due to a reverse external shock. One is that the
optical light curve does not show a good correlation to the gamma-ray
light curve (Sari \& Piran ,1999b), and the model fits generally well.
The other is that the flash from internal shocks is weaker than from
external reverse shocks, as discussed here. On the other hand, the
first ROTSE observations of GRB 990123 (Akerlof \etal, 1999a) started 20 
seconds after the GRB trigger, and do not appear to have sampled the optical 
light curve densely enough to establish the degree of correlation with
good significance (both curves are compared in Fenimore \etal, 1999).
While the likelihood of detecting an external reverse shock optical flash
appears to be higher at this stage, one needs a faster triggering and a better 
sampling of the optical data in order to discriminate between reverse and 
internal shocks. An interesting possibility is that, since the external reverse 
shock starts somewhat later than the internal shocks, one might initially see a 
weaker optical flash from the internal shocks, which are overtaken by a stronger 
reverse external shock radiation after tens of seconds, until the forward 
shock optical afterglow takes over after 300-1000 s.  Especially for the more 
frequent weaker bursts, the prospect of investigating such features underlines 
the need for dedicated GRB afterglow missions such as HETE2 and Swift.

\acknowledgments
This research is supported by NASA NAG5-2857,
NSF PHY94-07194 and the Royal Society. We are grateful to A. Panaitescu
for useful insights.

\end{document}